\pdfoutput=1

\documentclass[peerreview]{IEEEtran}
\usepackage{cite}

\usepackage{color}

\ifCLASSINFOpdf
\else
\fi
\usepackage{amsmath}
\usepackage{bm}
\usepackage{amsfonts}
\usepackage{mathtools}
\usepackage{algorithmic,algorithm2e,float}

\usepackage{array}
\usepackage{mathtools}
\begin{document}

\title{Coded convolution for parallel and distributed computing within a deadline}
%
\IEEEoverridecommandlockouts
\author{Sanghamitra~Dutta, Viveck~Cadambe and  Pulkit~Grover
\thanks{This paper is to appear in ISIT 2017. Sanghamitra Dutta and Pulkit Grover are with Department of Electrical and Computer Engineering, Carnegie Mellon University, Pittsburgh, PA 15213, USA. Viveck Cadambe is with the Department of Electrical Engineering at Pennsylvania State University, University Park, PA-16802, USA. }}

\maketitle

\begin{abstract} 
We consider the problem of computing the convolution of two long vectors using parallel processing units in the presence of ``stragglers''. Stragglers refer to the small fraction of faulty or slow processors that delays the entire computation in time-critical distributed systems. We first show that splitting the vectors into smaller pieces and using a linear code to encode these pieces provides better resilience against stragglers than replication-based schemes under a simple, worst-case straggler analysis. We then demonstrate that under commonly used models of computation time, coding can dramatically improve the probability of finishing the computation within a target ``deadline'' time. As opposed to the more commonly used technique of expected computation time analysis, we quantify the exponents of the probability of failure in the limit of large deadlines. Our exponent metric captures the probability of failing to finish before a specified deadline time, \textit{i.e.}, the behavior of the ``tail''. Moreover, our technique also allows for simple closed form expressions for more general models of computation time, \textit{e.g.} shifted Weibull models instead of  only shifted exponentials. Thus, through this problem of coded convolution, we establish the utility of a novel asymptotic failure exponent analysis for distributed systems.
\end{abstract}

\setlength{\parskip}{8pt}
\setlength{\parsep}{8pt}
\setlength{\partopsep}{8pt}


\section{Introduction}
\label{sec:introduction}
\IEEEPARstart{T}{he} operation of convolution has widespread applications in mathematics, physics, statistics and signal processing. Convolution plays a key role in solving inhomogeneous differential equations to determine the response of a system under initial conditions \cite{arfken2005mathematical}. Convolution of any arbitrary input function with the pre-specified impulse response of the system, also known as the Green's function \cite{arfken2005mathematical}, provides the response of the system to any arbitrary input function. Convolution  is also useful for filtering or extraction of features, in particular for Convolutional Neural Networks (CNNs), which are frequently used in time-critical systems like autonomous vehicles.  In \cite{dally2015}, Dally points out that application of machine learning tools in low-latency inference problems is primarily bottlenecked by the computation time.

In this paper, we propose a novel coded convolution for time-critical distributed systems prone to straggling and delays, for fast and reliable computation within a target deadline. We also propose a novel analysis technique based on a new metric -- the exponent of the asymptotic probability of failure to meet a target deadline -- for comparison of various delay tolerant techniques in distributed systems. Our contributions are two fold:-
\begin{enumerate}
  \item We go beyond distributed matrix-vector products, and explore the problem of distributed convolution (building on \cite{yang2016fault}) in a straggler prone time-critical distributed scenario. We propose a novel strategy of splitting the vectors to be convolved and coding them, so as to perform fast and reliable convolution within specified deadlines. Moreover our strategy also allows for encoding and decoding online as compared to existing work on coded matrix-vector products which require encoding to be performed as a pre-processing step because of its high computational complexity.
    \item We introduce a new analysis technique to compare the performance of various fault/delay tolerant strategies in distributed systems. Under a more generalized shifted Weibull\cite{faultbook} computation time model (that also encompasses shifted exponential), we demonstrate the utility of the proposed deadline-driven analysis technique for the coded convolution problem.
\end{enumerate}

The motivation for our work arises from the increasing drive towards distributed and parallel computing with ever-increasing data dimensions. Parallelization reduces the per processor computation time, as the task to be performed at each processor is substantially reduced.  However, the benefits of parallelization are often limited by ``stragglers'': a few slow processors that delay the entire computation. These processors can cause the probabilistic distribution of computation times to have a long tail, as pointed out in the influential paper of Dean and Barroso\cite{straggler_tail}. For time-critical applications, stragglers hinder the completion of tasks within a specified deadline, since one has to wait for all the processors to finish their individual computations. 

The use of replication\cite{gauristraggler}, \cite{gauri2014delay}, \cite{gauriefficient} is a natural strategy for dealing with stragglers in distributed systems since the computation only requires a subset of the processors to finish. Error correcting codes have been found to offer advantages over replication in various situations. The use of simple check-sum based codes for correcting errors in linear transforms dates back to the ideas of algorithmic fault tolerance \cite{ABFT1984} and its extensions in \cite{faultbook} \cite{ABFT2009}. A strategy of coding multiple parallel convolutions over reals for error tolerance is proposed in \cite{yang2016fault}. An alternative technique for error-resilient convolutions, that limits itself to finite-field vectors and requires coding both vectors into residue polynomials, was proposed in \cite{sundaram2008fault}. Recently, the use of erasure codes, \textit{e.g.},  MDS code based techniques \cite{kananspeeding}\cite{mohammad2016}, has been explored  for speeding up matrix-vector products in distributed systems.  In \cite{dutta2016short}, we introduce a novel class of codes called -- Short-Dot codes -- that compute multiple straggler-tolerant short dot products towards computing a large matrix-vector product. The most common metric for comparing the performance of various strategies in existing literature is expected computation time that often uses  a shifted exponential computation time model. However, this method of analysis becomes unwieldy when extended to higher order moments. Moreover, expected time does not capture the probability of failure to meet a specific deadline which could be important in time-critical applications.



\section{Assumptions and Problem Formulation}
\label{sec:problem formulation} 

Before proceeding further, note that the computational complexity of convolving  two vectors of any lengths, say $m_1$ and $m_2$ using Fast Fourier Transform (FFT)   is  $\Theta \left((m_1 + m_2 -1) ( \log{(m_1 + m_2-1)} +1 )\right)$ \cite{cooley1969fast} \cite{bracewell1999fourier}. When one of $m_1$ or $m_2$ is much smaller than the other, say $\log{m_1} \ll \log{m_2}$, then the computational complexity can be reduced even further using overlap methods like overlap add \cite{overlapadd}. When using overlap methods, the longer vector is divided into smaller pieces of length comparable to the smaller one and each piece is convolved separately. The outputs are then combined together. In this paper, for conceptual simplicity we make some assumptions on the computational complexity of convolution, under different scenarios.

\noindent \textbf{Scenario 1:} We assume that when $m_1$ and $m_2$ are comparable, the computational complexity of convolution using FFT is given by $C (m_1 + m_2) (\log{(m_1 + m_2)})$ where $C$ is a constant.

\noindent \textbf{Scenario 2:} We assume that when the length of one vector is sufficiently smaller than another, specifically $ \log(m_2) = o(\log(m_1))$, then the computational complexity using overlap methods\cite{overlapadd} is $2Cm_1 (\log{(2m_2)}+ 1)$ where $C$ is a constant.


\subsection*{Problem Formulation}
We consider the problem of convolution of a vector $\bm{a}$ of length $N_1$ with an unknown input vector $\bm{x}$ of length $N_2$, using $P$ parallel processing units. The computation goal is to convolve the two vectors reliably in a distributed and parallelized fashion, in the presence of stragglers, so as to prevent failure to meet a specified deadline. 

\noindent \textbf{Assumption:} For the ease of theoretical analysis, we assume that $2\sqrt{\frac{N_1 N_2}{P}} \leq \min\{N_1, N_2\} $. This ensures that $N_1$ and $N_2$ are not too far from each other for \textit{relatively} small $P$.

Note that, the operation of convolution can be represented as the multiplication of a Toeplitz matrix, with a vector, requiring a naive computational complexity of $\Theta(N_1 N_2)$. One might wonder if we should parallelize this matrix vector product and, in doing so, use techniques from \cite{kananspeeding}\cite{dutta2016short} that make matrix-vector products resilient to straggling. However, using such a parallelization scheme over the $P$ given processors, the computational complexity per processor cannot be reduced to below $\Theta(\frac{N_1N_2}{P})$. If $P$ is not large enough, this is still substantially larger than $C(N_1 +N_2)\log{(N_1 + N_2)}$, the computational complexity of convolution using FFT on a single processor under Scenario 1. 

\section{The Naive Uncoded Strategy}
\label{sec:uncoded}
\begin{figure}
    \centering
    \includegraphics[height=5cm]{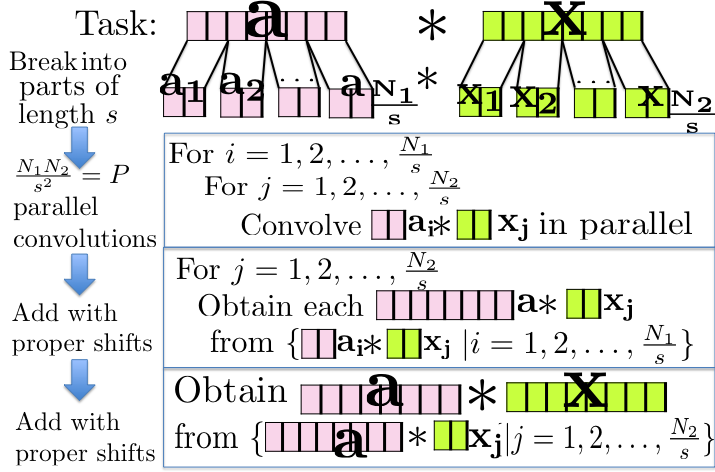}
    \caption{Naive Uncoded Convolution for two vectors of lengths $N_1$ and $N_2$ using $P$ parallel processors by dividing into pieces of length $s=\sqrt{N_1N_2/P}$.} \label{fig:uc_full}
\end{figure}
An uncoded strategy to perform the convolution is to divide both the vectors $\bm{a}$ (length $N_1$) and $\bm{x}$ (length $N_2$) into smaller parts of lengths $s_1$ and $s_2$ respectively. Let the parts of the vector $\bm{a}$ be denoted by $\bm{a}_1, \bm{a}_2,\dots,\bm{a}_{\frac{N_1}{s_1}} $, each of length $s_1$. Similarly, the parts of vector $\bm{x}$ are denoted by $\bm{x}_1,\bm{x}_2,\dots,\bm{x}_{\frac{N_2}{s_2}}$. We first consider the case where the entire convolution is computed in parallel using all the given $P$ processors, in one shot. Consider the following algorithm:\\

\begin{algorithmic}
\STATE \hspace{5cm} \textit{\textbf{For }}$i = 1,2,\dots,N_1/s_1$
 \STATE \hspace{6 cm} \textit{\textbf{For }}$j = 1,2,\dots,N_2/s_2$
 \STATE \hspace{7 cm} \textit{\textbf{Convolve }}$\bm{a}_i \ast \bm{x}_j$ \textbf{\textit{in parallel}}
\end{algorithmic}

\noindent Thus, we must have $\frac{N_1}{s_1} \frac{N_2}{s_2} = P$. Depending on the lengths $s_1$ and $s_2$, the convolutions can be performed under Scenario 1 or 2. First consider the case where we perform the convolutions under Scenario 1, choosing $s_1$ and $s_2$ of comparable lengths. Then, for minimizing per processor computational complexity, we have the following optimization problem:
\begin{align}
   & \min_{s_1, s_2} \   (s_1 + s_2) \log{(s_1 + s_2)}  \\
   & \text{subject to } s_1 s_2 = N_1 N_2/P \nonumber \\
   & 1 \leq s_1 \leq N_1 ,\  1 \leq s_2 \leq N_2 \nonumber
\end{align}
A quick differentiation reveals that the above minimization is achieved when $s_1= s_2= \sqrt{N_1 N_2/P}$ (say $s$). Thus, the computational complexity required for convolution under Scenario 1 would be $C\big(2\sqrt{ \frac{N_1 N_2}{P}}\big)\log{ \Big(2\sqrt{ \frac{N_1 N_2}{P}}\Big)}$. We show in Appendix A that operating under Scenario 2 requires equal or higher per processor complexity than the aforementioned strategy. We also show that, when we use the $P$ processors several times instead of operating in a single shot, the computational complexity only increases.

Thus, for uncoded convolution, we provably show that the minimum computational complexity per processor in order sense is attained when $s_1=s_2=\sqrt{N_1 N_2/P}$ (say $s=\sqrt{N_1 N_2/P}$). Now, we can perform $P$ convolutions of $s$-length vectors, convolving each of the $\frac{N_1}{s}$ parts of $\bm{a}$ with $\frac{N_2}{s}$ parts of $\bm{x}$ in parallel. A convolution of length $s=\sqrt{N_1N_2/P}$ in each processor has a computational complexity of $2C \sqrt{\frac{N_1 N_2}{P}} \log{2\sqrt{\frac{N_1 N_2}{P}}}$, where $C$ is a constant which is independent of the length of the convolution. Fig.~\ref{fig:uc_full} shows the uncoded convolution strategy.

\textbf{Reconstruction:} First let us discuss how to reconstruct each $\bm{a}\ast \bm{x}_j$ of length $N_1+s_2$ from successfully computed convolution outputs $\{\bm{a}_i \ast \bm{x}_j | i=1,2,\dots,N_1/s_1\}$ each of size $s_1 + s_2$. Consider the following algorithm:\\ 

\begin{algorithmic}
\STATE \hspace{5cm}\textbf{\textit{For }} $i= 1,2,\dots,N_1/s_1$
\STATE \hspace{6cm}\textbf{\textit{Pad }}$\bm{a}_i \ast \bm{x}_j$ \textbf{\textit{with}} $0$\textbf{\textit{s to the right to get length}} $N_1+s_2$
\STATE \hspace{6cm}\textbf{\textit{Shift}} $\bm{a}_i \ast \bm{x}_j$ \textbf{\textit{ right by }} $(i-1)s_1 $
 \STATE \hspace{5cm}\textbf{\textit{Finally add all }} $N_1/s_1$ \textbf{ \textit{shifted outputs}}
\end{algorithmic}

\begin{figure}
    \centering
    \includegraphics[height=3.4cm]{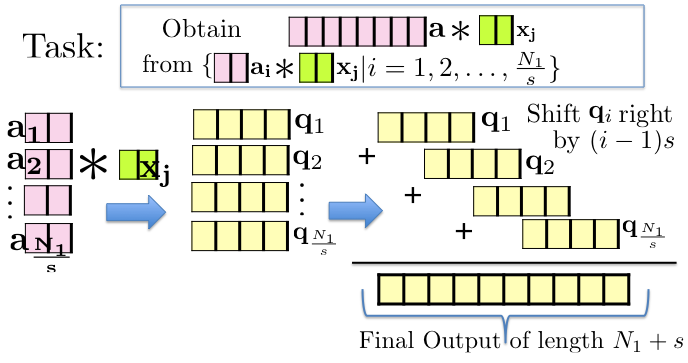}
    \caption{The process of combining outputs with appropriate shifts to generate $\bm{a}\ast \bm{x}_j$ from $\{\bm{a}_i \ast \bm{x}_j | i=1,2,\dots,N_1/s_1\}$, where $s_1=s_2=s$.}
    \label{fig:uc_part}
\end{figure}
The algorithm is also illustrated in  Fig.~\ref{fig:uc_part} for $s_1=s_2=s$. Now, we may similarly obtain $\bm{a}\ast \bm{x}$, from $\{\bm{a} \ast \bm{x}_j | j=1,2,\dots,N_2/s_2\}$, by shifting each result right by $(j-1)s_2$, and adding the shifted results all together.

Note that this uncoded strategy requires all the $P$ processors to finish. If some processors straggle, the entire computation is delayed, and may fail to finish within the specified deadline. In the next sections, we first analyze a replication strategy that introduces redundancy and does not require all the processors to finish, and then propose our coded convolution strategy.




\section{The Replication Strategy}
\label{sec:replication}
In the previous section, we have shown that the strategy of minimizing the per processor computational complexity in convolution  is to divide each vector into equal sized portions, and convolve in parallel processors, all in one shot. Now, let us consider a $(P,r)$ replication strategy, where every sub task has $r$ replicas or copies. We basically divide the task of convolution into $P/r$ equal parts, and each part has $r$ replicas, so as to use all the given $P$ processors. Then, the length of the vectors to be convolved at each processor is given by $s=\sqrt{\frac{N_1 N_2 r}{P}}$. 
\newtheorem{theorem}{Theorem}
{\color{black}{
\begin{theorem}
For the convolution of a vector of length $N_1$ with another vector of length $N_2$ using $P$ processors,  a $(P,r)$ replication strategy needs  $K=P-\frac{Ps^2}{N_1N_2}+1$ short convolutions of length $s=\sqrt{\frac{N_1 N_2 r}{P}}$ to finish in the worst case.
\end{theorem}
}}
\begin{IEEEproof}
Observe that there are $\frac{P}{r}$ tasks that need to finish, each having $r$ replicas. The worst case wait arises when the replicas of all but one of the $\frac{P}{r}$ tasks have completed before any one copy of the last task has finished. Thus,

\begin{equation}
K=\Big( \frac{P}{r} - 1 \Big) r + 1 = P-\frac{Ps^2}{N_1N_2}+1
\end{equation}
This proves the theorem.\end{IEEEproof}

\section{The Coded Convolution Strategy}
\label{sec:codedconv}
We now describe our proposed coded convolution strategy which ensures that computation outputs from a subset of the $P$ processors are sufficient to perform the overall convolution.

\subsection{An $(N_1,N_2,P,s)$ Coded Convolution} 
\begin{figure}
    \centering
    \includegraphics[height=6cm]{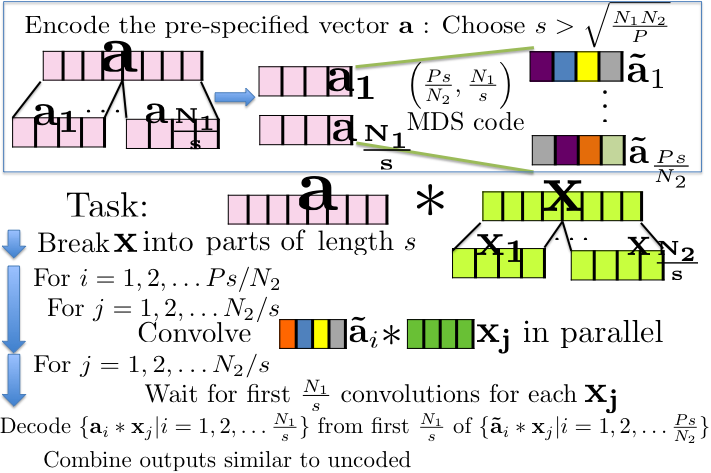}
    \caption{Coded Convolution strategy for two vectors of lengths $N_1$ and $N_2$ using $P$ parallel processors by dividing into pieces of length $s>\sqrt{N_1N_2/P}$.} \label{fig:codedconv}
\end{figure}

Let us choose a length $s$ such that $\sqrt{\frac{N_1 N_2}{P}} < s $. Now, we can divide both the vectors $\bm{a}$ and $\bm{x}$ into parts of length $s$, as shown in Fig.~\ref{fig:codedconv}. The vector $\bm{x}$ has $\frac{N_2}{s}$ parts of length $s$, namely $\bm{x}_1,\dots,\bm{x}_{\frac{N_2}{s}}$. Similar to uncoded strategy, for every convolution $\bm{a} \ast \bm{x}_j$  we assign $Ps/N_2$ processors .  The $\frac{N_1}{s}$ parts of the  vector $\bm{a}$ are encoded using a $\big(\frac{Ps}{N_2}, \frac{N_1}{s} \big)$ MDS code to produce $\frac{Ps}{N_2}$ vectors of length $s$, as discussed below: 

Let $\bm{G}_{ \big(\frac{N_1}{s} \times \frac{Ps}{N_2}\big) }$ be the generator matrix of a $\left(\frac{Ps}{N_2}, \frac{N_1}{s} \right)$ MDS code on the real field, \textit{i.e.}, a matrix such that every $\frac{N_1}{s} \times \frac{N_1}{s}$ square sub-matrix is invertible \cite{dutta2016short}. For example, an $\big(  \frac{N_1}{s} \times \frac{Ps}{N_2} \big)$ Vandermonde matrix satisfies this property.  Now we encode the $\bm{a}_i$s as follows:
\begin{equation}
\bm{G}^T \begin{bmatrix}
\bm{a}_1 \\
\vdots\\
\bm{a}_{\frac{N_1}{s}}
\end{bmatrix}= \begin{bmatrix}
1 & g_{1}  & \dots & g_{1}^{\frac{N_1}{s} -1 }\\
1 &\vdots  & \ddots  & \vdots \\
1 & g_{\frac{Ps}{N_2}} & \dots & g_{\frac{Ps}{N_2}}^{\frac{N_1}{s} -1 }
\end{bmatrix}\begin{bmatrix}
\bm{a}_1 \\
\vdots\\
\bm{a}_{\frac{N_1}{s}}
\end{bmatrix}=
\begin{bmatrix}
\bm{\tilde{a}}_1 \\
\vdots\\
\bm{\tilde{a}}_{\frac{Ps}{N_2}}
\end{bmatrix}
\end{equation}

Here $\{g_1,g_2,\dots, g_{\frac{Ps}{N_2}} \}$ are all distinct reals. Since every $\frac{N_1}{s} \times \frac{N_1}{s}$ square sub-matrix of $\bm{G}$ is invertible, any $\frac{N_1}{s}$ vectors out of the $\frac{Ps}{N_2}$ coded vectors $\{\bm{\tilde{a}}_1,\dots,\bm{\tilde{a}}_{\frac{Ps}{N_2}}\}$ can be linearly combined to generate the original $\frac{N_1}{s}$ parts of $\bm{a}$, \textit{i.e.}, $\{\bm{a}_1,\dots,\bm{a}_{\frac{N_1}{s}}\}$. To see this, let $L=\{l_{1},l_{2},\dots, l_{\frac{N_1}{s}}\} $ denote a subset of any $\frac{N_1}{s}$ distinct indices of $ \{1,2,\dots,\frac{Ps}{N_2}\}$. Also, let $\{\bm{\tilde{a}}_{l_{1}},\dots,\bm{\tilde{a}}_{l_{\frac{N_1}{s}}}\}$ be the set of $\frac{N_1}{s}$ coded vectors corresponding to the indices in $L$. Thus,
$$
\bm{G}^T_L \begin{bmatrix}
\bm{a}_1 \\
\vdots\\
\bm{a}_{\frac{N_1}{s}}
\end{bmatrix}= \begin{bmatrix}
\bm{\tilde{a}}_{l_{1}} \\
\vdots\\
\bm{\tilde{a}}_{l_{\frac{N_1}{s}}}
\end{bmatrix}.
$$ Here $\bm{G}^T_L$ denotes a $\frac{N_1}{s} \times \frac{N_1}{s}$ square sub-matrix of $\bm{G}^T$, consisting of the rows indexed in $L$. Since $\bm{G}^T_L$  is invertible, we have
\begin{equation}
 \begin{bmatrix}
\bm{a}_1 \\
\vdots\\
\bm{a}_{\frac{N_1}{s}}
\end{bmatrix}= [\bm{G}^T_L]^{-1}\begin{bmatrix}
\bm{\tilde{a}}_{l_{1}} \\
\vdots\\
\bm{\tilde{a}}_{l_{\frac{N_1}{s}}}
\end{bmatrix} = \bm{B} \begin{bmatrix}
\bm{\tilde{a}}_{l_{1}} \\
\vdots\\
\bm{\tilde{a}}_{l_{\frac{N_1}{s}}}
\end{bmatrix} \label{decoding}
\end{equation}
Here, $\bm{B}$ denotes $[\bm{G}^T_L]^{-1}$. Now, each of these coded vectors $\{\bm{\tilde{a}}_1,\dots,\bm{\tilde{a}}_{\frac{Ps}{N_2}}\}$ is convolved with each part of $\bm{x}$, \textit{i.e.}, $\{\bm{x}_1,\dots,\bm{x}_{\frac{N_2}{s}} \}$ in $P$ separate processors. Fig.~\ref{fig:codedconv} shows the coded convolution strategy.
From \eqref{decoding}, for any $j$, all the original $\{\bm{a}_i \ast \bm{x}_j | i=1,2,\dots,\frac{N_1}{s}\}$ can be decoded from a linear combination of any $\frac{N_1}{s}$ of $\{ \bm{\tilde{a}}_1 \ast \bm{x}_j, \bm{\tilde{a}}_2 \ast \bm{x}_j, \dots, \bm{\tilde{a}}_{\frac{Ps}{N_1}} \ast \bm{x}_j \}$, in particular, those that finish first. Let $\bm{B}_{i}$ denote a row vector corresponding to the $i^{th}$ row of $\bm{B}$. Thus for any $(i,j)$,
\begin{equation}
    \bm{a}_i \ast \bm{x}_j= \bm{B}_{i} \begin{bmatrix}
\bm{\tilde{a}}_{l_{1}} \ast \bm{x}_j \\
\vdots\\
\bm{\tilde{a}}_{l_{\frac{N_1}{s}}} \ast \bm{x}_j
\end{bmatrix} \Rightarrow \begin{bmatrix}
\bm{a}_1 \ast \bm{x}_j \\
\vdots\\
\bm{a}_{\frac{N_1}{s}} \ast \bm{x}_j
\end{bmatrix}= \bm{B} \begin{bmatrix}
\bm{\tilde{a}}_{l_{1}} \ast \bm{x}_j\\
\vdots\\
\bm{\tilde{a}}_{l_{\frac{N_1}{s}}} \ast \bm{x}_j
\end{bmatrix} \nonumber
\end{equation}
The decoded outputs can be combined similar to the uncoded strategy. Now, we state some results for Coded Convolution.{\color{black}{
\begin{theorem} 
For convolution of a vector of length $N_1$ with another vector of length $N_2$ using $P$ processors, an  $(N_1,N_2,P,s)$ Coded Convolution decoder needs  $K=P -\frac{Ps}{N_2} + \frac{N_1}{s}$ short convolutions of length $s=\sqrt{\frac{N_1N_2}{P}}$ to finish in the worst case.
\end{theorem}  }}
\begin{IEEEproof}
Note that each part of $\bm{x}$ of length $s$ is convolved with $\frac{Ps}{N_2}$ vectors, so that any $\frac{N_1}{s}$ are sufficient. Now, in the worst case, all but one part of $\bm{x}$ finishes convolution with all $\frac{Ps}{N_2}$ vectors before the last one finishes with $\frac{N_1}{s}$ vectors. Thus, the number of processors required is at most,

\begin{equation}
K= \Big(\frac{N_2}{s}-1\Big)\frac{Ps}{N_2} + \frac{N_1}{s} = P -\frac{Ps}{N_2} + \frac{N_1}{s}
\end{equation}
This proves the theorem.
\end{IEEEproof}

 \subsection{Comparison with replication strategy:} We first show that the worst-case number of processors required using Coded Convolution is less than that required using replication for the same per-processor computational complexity.
\newtheorem{corollary}{Corollary}
\begin{corollary}
For convolution of vector of length $N_1$ with an unknown input vector of length $N_2$ using $P$ processors, an $(N_1,N_2,P,s)$ Coded Convolution decoder waits for fewer processors in the worst case as compared to that of a $(P,r)$ replication strategy with equal per processor computational complexity.
\end{corollary}

\begin{IEEEproof}
Recall from Section \ref{sec:replication} that in a $(P,r)$ replication strategy, the length of the vectors convolved at each processor is given by $\sqrt{\frac{N_1 N_2 r}{P}}$. For equal per processor task, this length should be equal to $s$ in Coded Convolution. Using the fact that $s= \sqrt{\frac{N_1 N_2 r}{P}} > \sqrt{\frac{N_1 N_2}{P}}$, we get,
\begin{equation}
K_{Coded}= P -\frac{Ps}{N_2} + \frac{N_1}{s} < P-\frac{Ps^2}{N_1N_2}+1 = K_{replication}
\end{equation}
This proves the theorem.
\end{IEEEproof}

 \noindent \textbf{Comment:} To justify our focus on the worst case $K$, we also discuss in Section \ref{sec:asymptotics} how the exponent of the probability of failure to meet a deadline in the asymptotic regime depends on the worst case $K$.

\subsection{Computational Complexity of Encoding and Reconstruction:} Observe that, the computational complexity of each processor is $(2Cs)\log{(2s)}$ from Scenario 1. To be able to encode and reconstruct online, it is desirable that the encoding complexity and reconstruction complexity, \textit{i.e.}, the total complexity of \textbf{decoding and subsequent additions} are both negligible compared to the per processor complexity, since otherwise the encoder/decoder becomes the primary bottleneck during successful completion of tasks within a specified deadline.

\begin{theorem}
\label{thm:decoding}
For an $(N_1,N_2,P,s)$ Coded Convolution using a Vandermonde Encoding Matrix, the ratio of the total encoding complexity and reconstruction complexity to the per processor complexity tends to $0$, as $N_1,N_2,P \to \infty$, when
$ P (\log{P})^2 = o \Big( \log{\sqrt{\frac{N_1 N_2}{P}}} \Big)
$.
\end{theorem} 
\noindent A proof based on \cite{kung1973fast}, \cite{li2000arithmetic} is provided in Appendix B.

\section{Asymptotic Analysis of Exponents}
\label{sec:asymptotics}
We now perform an asymptotic analysis of the exponent of the probability of failure to meet a deadline in the limit of the deadline diverging to infinity. Let $F_s(t)$ denote the probability that a convolution of two vectors of length $s$ is computed in a single processor within time $t$, based on a shifted Weibull model. Thus   
\begin{equation}
F_s(t)= 1-e^{-\big(\mu \left( \frac{t}{2Cs\log{(2s)}}-1\right) \big)^\alpha}\ \ \forall\  t \ \geq  2Cs\log{(2s)}
\end{equation}

\noindent Here $\mu > 0$ is a straggling parameter, $\alpha > 0$ denotes the exponent of the delay distribution and $C$ denotes the constant of convolution that is independent of the length of the vectors. For shifted exponential model, $\alpha=1$.

Let $P^f_s(t)$ be the probability of failure to finish the convolution of two vectors of length $s$ at or before time $t$.

From Theorem \ref{thm:decoding}, when $P (\log{P})^2 = o \big( \log{\sqrt{N_1N_2/P}} \big)$, the decoding complexity is negligible compared to the per processor complexity. Thus, only the straggling in the parallel processors is the bottleneck for failure to meet a deadline. We now compute the failure-exponent for the coded strategy, in the limit of $t \to \infty$. For the coded strategy, at most $K=P-\frac{P s}{N_2}+\frac{N_1}{s}  $ processors need to finish computation in the worst case. An \textbf{upper bound} on the failure probability is thus given by the probability that at most $K-1$ processors have finished the convolution of two $s$ length vectors. Thus,
\begin{align}
P^f_s(t)  &\leq \sum_{i=0}^{K-1} \binom{P}{i} (F_s(t))^i (1-F_s(t))^{P-i} \\ 
& = \sum_{i=0}^{K-1} \binom{P}{i} \Big(\frac{(F_s(t))}{1-F_s(t)}\Big)^i (1-F_s(t))^{P} \\
& =\Theta \Big(c(P,K) \Big(\frac{F_s(t)}{1-F_s(t)}\Big)^{K-1} (1-F_s(t))^{P} \Big)
\end{align}
Here $c(P,K)$ denotes a function of $P$ and $K$, that is independent of $t$. The last line follows since for $t$ large enough, $\frac{(F_s(t))}{(1-F_s(t))} \gg 1$, and thus for the purpose of analysis of the failure exponent for large $t$, the binomial summation is of the same order as the largest term, as we show here:
\begin{align}
 \binom{P}{K-1} \Big(\frac{(F_s(t))}{1-F_s(t)}\Big)^{K-1} (1-F_s(t))^{P} \leq \sum_{i=0}^{K-1} \binom{P}{i} \Big(\frac{(F_s(t))}{1-F_s(t)}\Big)^i (1-F_s(t))^{P}  \leq \sum_{i=0}^{K-1} \binom{P}{i} \Big(\frac{(F_s(t))}{1-F_s(t)}\Big)^{K-1} (1-F_s(t))^{P}
\end{align}

This implies,
\begin{multline}
 c_1(P,K)\Big(\frac{(F_s(t))}{1-F_s(t)}\Big)^{K-1} (1-F_s(t))^{P}  \leq \sum_{i=0}^{K-1} \binom{P}{i} \Big(\frac{(F_s(t))}{1-F_s(t)}\Big)^i (1-F_s(t))^{P}  \leq c_2(P,K)\Big(\frac{(F_s(t))}{1-F_s(t)}\Big)^{K-1} (1-F_s(t))^{P}
\end{multline}

\noindent Bounding the leading co-efficient in  the failure exponent, 
\begin{align}
 \lim_{t \to \infty}  \frac{\log(P^f_s(t))}{t^\alpha} 
& \leq \lim_{t\to \infty} 
\frac{\log \Big(\frac{c_2(P,K)}{1} \big(\frac{F_s(t)}{1-F_s(t)} \big)^{K-1} \frac{(1-F_s(t))^P}{1}\Big)} {t^\alpha}   \nonumber \\
& = \lim_{t\to \infty}  (K-1)\frac{\log \left( e^{\mu^\alpha \left( \frac{t}{(2Cs)\log{(2s)}}-1\right)^\alpha }-1\right)}{t^\alpha}  + P \frac{\log \left(e^{-\mu^\alpha \left( \frac{t}{(2Cs)\log{(2s)}}-1\right)^\alpha }\right)}{t^\alpha} \nonumber \\
& \to -\frac{(P-K+1)\mu^\alpha}{((2Cs)\log{(2s)})^\alpha} \nonumber \\
& = -\frac{\big( \frac{Ps}{N_2} -\frac{N_1}{s}  +1\big)\mu^\alpha}{((2Cs)\log{(2s)})^\alpha}
\label{exponent}
\end{align}.\\
\noindent \textit{Definition:} Let us define a function $\epsilon(s)$ as follows: $$\epsilon(s)= -\frac{\big( \frac{Ps}{N_2} -\frac{N_1}{s}  +1\big)\mu^\alpha}{((2Cs)\log{(2s)})^\alpha} $$ Note that $\epsilon(s)$ is an \textbf{upper bound} on the leading co-efficient of the failure exponent for an $(N_1,N_2,P,s)$ Coded Convolution. Now we compare coding with the uncoded strategy.

For the uncoded strategy in Section \ref{sec:uncoded}, the failure event occurs \textbf{exactly} (not upper bound)  when at most $P-1$ processors have failed to finish. The failure exponent can thus be computed \textbf{exactly} using steps similar to \eqref{exponent}. For the uncoded strategy, the length of the vectors in each processor is $s=\sqrt{N_1N_2/P}$ and the number of processors to wait for is $P$ (also given by $K = P-\frac{Ps}{N_2} +\frac{N_1}{s} $ from Theorem 2). The leading co-efficient in the failure exponent is exactly given by, 
\begin{equation} 
\label{uncoded exponent}
 \epsilon(s)|_{s=\sqrt{N_1N_2/P}}= -\frac{\mu^\alpha}{\Big(2C\sqrt{\frac{N_1N_2}{P}}\log{2\sqrt{\frac{N_1N_2}{P}}}\Big)^\alpha}
\end{equation}

\begin{theorem}
For $N_1=\Theta(N_2)$ and fixed Weibull parameter $\alpha$, there exists an $( N_1 , N_2, P, s ) $ Coded Convolution strategy with $s$ strictly greater than $\sqrt{N_1N_2/P}$ such that the ratio of the exponents of the asymptotic failure probability of the coded strategy to the naive uncoded strategy scales as $\Omega(\sqrt{P})$ and thus diverges to infinity for large $P$.
\end{theorem}

\begin{IEEEproof}
From \eqref{uncoded exponent}, the leading co-efficient in the failure exponent for the uncoded strategy is $ \epsilon\big(\sqrt{N_1N_2/P} \big)$. Now let us choose a coded strategy with $s=2\sqrt{N_1N_2/P}$. Putting this value of $s$ in $\epsilon(s)$, we get $\epsilon(s)=-\frac{\mu^\alpha(\frac{3}{2}\sqrt{\frac{PN_1}{N_2}}+1)}{ \left(4C\sqrt{\frac{N_1 N_2}{P}}\log{4\sqrt{\frac{N_1 N_2}{P}}}\right)^\alpha} $. Thus comparing the ratio,
\begin{equation}
\frac{\big|\epsilon\left(2\sqrt{N_1N_2/P} \right)\big|}{\left|\epsilon\left(\sqrt{N_1N_2/P} \right) \right|}= \frac{\frac{3}{2}\sqrt{\frac{PN_1}{N_2}}+1}{2^\alpha \Big( 1+ \frac{\log{2}}{\log{2\sqrt{N_1N_2/P}}}  \Big)^\alpha } > \frac{3.\sqrt{\frac{PN_1}{N_2}}}{2.4^\alpha} 
\end{equation}
Since from the conditions of the theorem, $N_1=\Theta(N_2)$, the ratio scales as $\Omega(\sqrt{P})$ and diverges for large $P$.
\end{IEEEproof}

\noindent \textbf{Comment:} The best choice of $s$ would be an integer in the range $\big(\sqrt{N_1N_2/P}, \min\{N_1,N_2\} \big]$ that maximizes $\left|  \epsilon(s) \right| $.

\subsection*{Comparison with Replication:} 
\begin{figure}[!t]
\centering
\includegraphics[height=4cm]{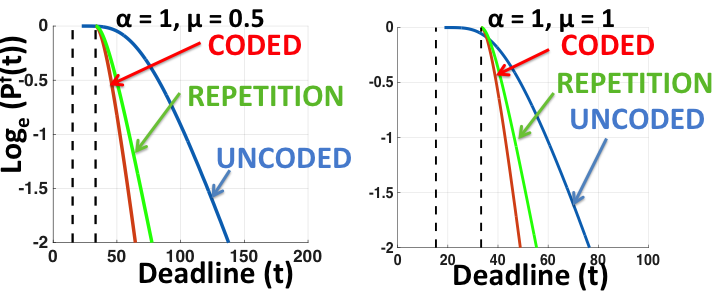}
\caption{Simulation Results: The plot shows the log of the complement of the cdf, \textit{i.e.} ($1-$cdf) of the computation time for an uncoded strategy, an $(2^{12},2^{11},8,2^{11})$ Coded Convolution strategy and a $(8,4)$ Repetition strategy based on $10^6$ Monte Carlo simulations from their respective shifted exponential distributions in MATLAB (Code available in \cite{duttaconv}). We observe that for uncoded strategy the decay of the failure exponent starts first, but is outperformed by both repetition and coded convolution as the deadline becomes large due to steeper rate of decay. Coded convolution is found to have the steepest decay for large deadlines. } 
\label{fig:simulation}
\end{figure}

We show in Appendix C that the exponent of the probability of failure to meet a deadline in the asymptotic regime depends on the worst case $K$ for replication. In this paper, we include simulations in Fig.~\ref{fig:simulation} showing that coded convolution outperforms replication in terms of the exponent of the probability of failure to meet a deadline. Moreover the exponents observed from the simulations are quite close to those calculated theoretically using the worst case $K$.

\noindent \textbf{Simulation Results:}
 We consider the convolution of a given vector of length $N_1=2^{12}$ with an unknown vector of length $N_2=2^{11}$, using $P=8$ parallel processors, based on an exponential model in MATLAB. Our results in Fig.~\ref{fig:simulation} show that coded convolution has the fastest decay of failure exponent for large deadlines. The slopes obtained from simulations are found to be quite close to the theoretically calculated values.

\section{Discussion} Thus, the proposed strategy outperforms existing strategies in terms of the exponent of the probability of failure to meet a deadline, in the limit of large deadlines. It might be observed that our strategy allows for online encoding and reconstruction as compared to existing works on coded matrix-vector products where the encoding cost can be amortized only if the matrix is pre-specified.



 
\section*{Acknowledgment}

This work was supported in part by Systems on Nanoscale Information fabriCs (SONIC), one of the six SRC STARnet Centers, sponsored by MARCO and DARPA. The support of NSF Awards 1350314, 1464336 and 1553248 is also acknowledged. Sanghamitra Dutta also received Prabhu and Poonam Goel Graduate Fellowship.

The authors would like to thank Franz Franchetti, Tze Meng Low and Doru Thom Popovici for their useful feedback regarding this research. Praveen Venkatesh, Haewon Jeong and Yaoqing Yang are also thanked for their comments and suggestions.

\ifCLASSOPTIONcaptionsoff
  \newpage
\fi



\bibliographystyle{IEEEtran}
\bibliography{IEEEabrv,sample}

\appendices
\section{Optimal Uncoded Strategy}
\label{appendix1}

We prove that performing convolution under Scenario 2, results in equal or higher per processor complexity. Without loss of generality, assume that $s_2 \leq s_1$. Then, under Scenario 2, the complexity of the convolution is given by  $C (2s_1) ( \log{(2s_2)}+1) $. Now, we are to solve
\begin{align}
&\min_{s_1,s_2} C (2s_1) (\log{(2s_2)}+1) \\
&\text{subject to } s_1 s_2 = \frac{N_1 N_2}{P} \\
& 1\leq s_1 \leq N_1 ,\ 1 \leq s_2 \leq N_2 
\end{align}

Examine the function
\begin{equation}
f(s_1)=C (2s_1) \left( \log{\left(\frac{2N_1 N_2}{Ps_1}\right)}+1 \right) 
\end{equation}

This function is concave in $s_1$. Let us examine its derivative.
\begin{equation}
f'(s_1)=2C\left(\log{\frac{2N_1N_2}{P}} -\log{s_1}\right)
\end{equation}
The derivative is positive in the range of $s_1$, \textit{i.e.} $1\leq s_1 \leq \min \{N_1, \frac{N_1 N_2}{P}\}$. Thus, the minimum is attained for the least value of $s_1$. However, as the product of $s_1$ and $s_2$ is constant, $s_1$ has to be greater or equal to the geometric mean, since otherwise $s_2 > s_1$ which violates our assumption. Thus, the minimum value of $s_1$ is $\sqrt{\frac{N_1 N_2}{P}}$, which results in the same time complexity as convolution without overlap methods.

Now, we consider performing convolutions using several serial uses of the given $P$ parallel processors, say $r$ uses. Then, effectively we have $rP$ processors, and the length of the convolution vector in each serial use of a single processor is $\sqrt{\frac{N_1N_2}{rP}}$. As this operation is to be done $r$ times, the computational complexity may be written as $r (2\sqrt{\frac{N_1N_2}{rP}}) \log{(2\sqrt{\frac{N_1N_2}{rP}})} $, which increases roughly with a factor of $\sqrt{r}$, compared to convolution in a single shot. Thus, use of the processors serially, only worsens the computational complexity per processor, while still requiring all of them to finish.

\section{Encoding and Decoding Complexity}
\textit{Theorem 3:}
For an $(N_1,N_2,P,s)$ Coded Convolution using a Vandermonde Encoding Matrix, the ratio of the total encoding complexity and reconstruction complexity to the per processor complexity tends to $0$, as $N_1,N_2,P \to \infty$, when
$ P (\log{P})^2 = o \Big( \log{\sqrt{\frac{N_1 N_2}{P}}} \Big)
$.

\begin{IEEEproof}
\textit{Encoding}
For encoding, the first vector $\bm{a}$ is split into smaller pieces of length $s$, as given by $\{\bm{a}_1,\bm{a}_2,\dots,\bm{a}_{\frac{N_1}{s}} \}$. Now we encode the $\bm{a}_i$s as follows:
$$
\bm{G}^T \begin{bmatrix}
\bm{a}_1 \\
\vdots\\
\bm{a}_{\frac{N_1}{s}}
\end{bmatrix}= \begin{bmatrix}
1 & g_{1}  & \dots & g_{1}^{\frac{N_1}{s} -1 }\\
1 &\vdots  & \ddots  & \vdots \\
1 & g_{\frac{Ps}{N_2}} & \dots & g_{\frac{Ps}{N_2}}^{\frac{N_1}{s} -1 }
\end{bmatrix}\begin{bmatrix}
\bm{a}_1 \\
\vdots\\
\bm{a}_{\frac{N_1}{s}}
\end{bmatrix}=
\begin{bmatrix}
\bm{\tilde{a}}_1 \\
\vdots\\
\bm{\tilde{a}}_{\frac{Ps}{N_2}}
\end{bmatrix}
$$
Here $\bm{G}_{\frac{N_1}{s}, \frac{Ps}{N_2}}$ is chosen as a Vandermonde matrix and $\{g_1,g_2,\dots, g_{\frac{Ps}{N_2}} \}$ are all distinct reals. Observe that the encoding process thus becomes the evaluation of polynomials of degree $\frac{N_1}{s}-1$( Note that $\frac{N_1}{s}< \frac{Ps}{N_2}$)  at $\frac{Ps}{N_2}$ points, repeated $s$ times for $s$ length vector $\bm{a}_i$s. From \cite{kung1973fast}, \cite{li2000arithmetic}, it is known that the evaluation of a polynomial of degree $\frac{Ps}{N_2}-1$ at $\frac{Ps}{N_2}$ arbitrary points can be performed in $\mathcal
{O}(\frac{Ps}{N_2} (\log{\frac{Ps}{N_2}})^2)$. As this process is repeated $s$ times, the total encoding complexity is given by  $\mathcal{O}\left(s\left(\frac{Ps}{N_2}\right) (\log{\frac{Ps}{N_2} })^2\right)$.
Now observe that,
\begin{equation}
 (s)\left(\frac{Ps}{N_2}\right) \left(\log{\frac{Ps}{N_2}}\right)^2   
 < s P (\log{P})^2 \label{enc:ineq1}
 \end{equation}
Note that, \eqref{enc:ineq1} holds from the choice of $\sqrt{\frac{N_1 N_2}{P}} \leq s  $ which implies $ \frac{Ps}{N_2} \leq P$. Thus, the encoding complexity is $\mathcal{O}(s P (\log{P})^2)$.

\textit{Reconstruction}
The reconstruction complexity consists of the complexity of decoding and subsequent additions. First we compute the complexity of decoding.

For decoding, convolution outputs of length $2s$ arrive from each parallel processor. For reconstructing each of the $\frac{N_2}{s}$ parts of vector $\bm{x}$, any $\frac{N_1}{s}$ convolution outputs are sufficient. 
Recall that for any $\bm{x}_j$,
\begin{equation}
 \begin{bmatrix}
\bm{a}_1 \ast \bm{x}_j \\
\vdots\\
\bm{a}_{\frac{N_1}{s}} \ast \bm{x}_j
\end{bmatrix}= [\bm{G}^T_L]^{-1}\begin{bmatrix}
\bm{\tilde{a}}_{l_{1}} \ast \bm{x}_j\\
\vdots\\
\bm{\tilde{a}}_{l_{\frac{N_1}{s}}} \ast \bm{x}_j
\end{bmatrix}
\end{equation}

If $\bm{G}$ is chosen as a Vandermonde Matrix, the reconstruction problem takes the form,
\begin{equation}
 \begin{bmatrix}
\bm{a}_1 \ast \bm{x}_j \\
\vdots\\
\bm{a}_{\frac{N_1}{s}} \ast \bm{x}_j
\end{bmatrix}= \begin{bmatrix}
1 & g_{l_{1}}  & \dots & g_{l_{1}}^{\frac{N_1}{s} -1 }\\
1 & g_{l_{2}}  & \dots & g_{l_{2}}^{\frac{N_1}{s} -1 }\\  &
\vdots  & \ddots  & \vdots \\
1 & g_{l_{\frac{N_1}{s}}}  & \dots & g_{l_{\frac{N_1}{s}}}^{\frac{N_1}{s} -1 }
\end{bmatrix}^{-1}
\begin{bmatrix}
\bm{\tilde{a}}_{l_{1}} \ast \bm{x}_j\\
\vdots\\
\bm{\tilde{a}}_{l_{\frac{N_1}{s}}} \ast \bm{x}_j
\end{bmatrix}
\end{equation}
The reconstruction of $\{ \bm{a}_i \ast \bm{x}_j  \}$ for $i=1,2,\dots ,\frac{N_1}{s}$ thus reduces to the interpolation of a polynomial of degree $\frac{N_1}{s}-1$ from its value at $\frac{N_1}{s}$ known, arbitrary points, to be repeated $2s$ times, which is the length of the convolution output. From \cite{kung1973fast}, \cite{li2000arithmetic}, it is known that the interpolation of a polynomial of degree $\frac{N_1}{s}-1$ from its value at $\frac{N_1}{s}$ arbitrary points can be performed in $\mathcal
{O}(\frac{N_1}{s} (\log{\frac{N_1}{s}})^2)$. As this step is repeated $2s$ times for $\frac{N_2}{s}$ parts of vector $\bm{x}$, the total decoding complexity is given by $\mathcal{O}\left((2s)\left(\frac{N_1N_2}{s^2}\right) (\log{\frac{N_1}{s}})^2\right)$. Now observe that,
\begin{equation}
 (2s)\left(\frac{N_1 N_2}{s^2}\right) \left(\log{\frac{N_1}{s}}\right)^2   
 < 2s P (\log{P})^2 \label{ineq1}
 \end{equation}
Note that, \eqref{ineq1} holds from the choice of $\sqrt{\frac{N_1 N_2}{P}} \leq s  $ which also implies $ \frac{N_1}{s} \leq \frac{N_1 N_2}{s^2} \leq P$.

Now, let us consider the complexity of the subsequent additions. Note that, we have a total of $\frac{N_1}{s} \frac{N_2}{s}$ vectors to be added with appropriate shifts.  As each vector is of length $2s$, the complexity of adding any one vector is $\mathcal{O} (2s) $. Thus, the total complexity of the subsequent additions is given by $\mathcal{O}\big( (2s)\frac{N_1}{s} \frac{N_2}{s} \big)   $. Now observe that,
\begin{equation}
(2s)\frac{N_1}{s} \frac{N_2}{s} \leq 2sP \label{ineq2}
\end{equation}
Note that, \eqref{ineq2} also holds from the choice of $\sqrt{\frac{N_1 N_2}{P}} \leq s  $.

Thus, from \eqref{enc:ineq1}, \eqref{ineq1} and \eqref{ineq2}, we can bound the total encoding and reconstruction complexity as follows:
 \begin{align}
 &\text{Enc.} + \text{Dec.} + \text{Add.} \nonumber \\ & \leq \ \   D(s P (\log{P})^2 + s P (\log{P})^2 +
s P) \\
& =  o \left(2s\log{(2s)}\right) \label{rec_complexity}
\end{align}
Here $D$ is a constant. And \eqref{rec_complexity} follows from the conditions of the theorem that $P (\log{P})^2 = o \left( 2s \log{\sqrt{\frac{N_1 N_2}{P}}} \right)$ and from the choice of $\sqrt{\frac{N_1 N_2}{P}} \leq s  $.  Thus the total reconstruction complexity, \textit{i.e.}, the complexity of decoding and subsequent additions is $o \left(2s\log{(2s)}\right)$. Recall that the per processor complexity for performing convolutions of two vectors of length $s$ is given by $C (2s) \log{(2s)}$. Thus, the ratio of the decoding complexity to the per-processor time complexity tends to $0$ as $N_1, N_2, P \to \infty $.
\end{IEEEproof}

\section{Exponent for Repetition}
In Section \ref{sec:asymptotics}, we perform an asymptotic analysis of the exponent of the probability of failure to meet a deadline in the limit of the deadline diverging to infinity. Recall from Section \ref{sec:asymptotics} that an upper bound for the leading coefficient of the failure exponent for Coded Convolution is given by
\begin{equation}
\epsilon(s)= -\frac{( P -K  +1)\mu^\alpha}{((2Cs)\log{(2s)})^\alpha} 
\end{equation}

Here $K$ is the number of processors one needs to wait for in the worst case for Coded Convolution. We also showed that for uncoded strategy, this upper bound is actually exactly equal to the leading co-efficient of the failure exponent (putting worst case $K = P$) as the deadline diverges to infinity.
Now, we show that even for repetition strategy this upper bound is actually exactly equal to the upper bound using worst case $K$ as the deadline diverges to infinity.

\textit{Theorem 4:} For a $(P,r)$ repetition strategy, the leading coefficient of the exponent in the probability of failure to meet a deadline converges as
$$
\lim_{t \to \infty}  \frac{\log(P^f_s(t))}{t^\alpha} = \epsilon(s)= -\frac{( P -K  +1)\mu^\alpha}{((2Cs)\log{(2s)})^\alpha}
$$
where $K=P-r+1$ is the number of processors to wait for in the worst case for a $(P,r)$ repetition strategy.
\begin{IEEEproof}
Recall that in a $(P,r)$ repetition strategy, we basically divide the operation of convolution into $P/r$ equal tasks, and each task has $r$ repetitions, so as to use all the given $P$ processors. Let us introduce the notation $T_{i,j}$ to represent the random variable corresponding to the different computation times on different processors. Here $i$ is the task index which varies from $1$ to $P/r$ for the $P/r$ different parts of the convolution task. Also let $j$ be the replica index for each task which varies from $1$ to $r$ for $r$ replicas of each task. Thus $T_{i,j}$ represents the computation time of the $j$-th copy of the $i$-th task.

Consider any task with index $i$. Let $T'_{i}$ denote the completion time of the particular task. The task finishes when any one of its $r$ replicas finish. Thus, the distribution of $T'_{i}$ is given by
$$T'_{i}= \min_j\{T_{i,j}\}$$

Now let us try to find the distribution of $T'_{i}$. Recall that each $T_{i,j}$ is a random variable denoting the time required to perform a convolution of length $s$ at a processor where $s=\sqrt{\frac{N_1 N_2 r}{P}}$. Its computational complexity is given by $2Cs \log(2s)$.  From shifted Weibull model assumptions, the \textit{c.d.f.} of each $T_{i,j}$ is given by 
\begin{equation}
\Pr(T_{i,j}\leq t)=F_s(t)= 1-e^{-\big(\mu \left( \frac{t}{2Cs\log{(2s)}}-1\right) \big)^\alpha}\  \forall\  t \ \geq  2Cs\log{(2s)}
\end{equation}

Since the $\{T_{i,j}\}$s are \textit{i.i.d.}, we thus have
\begin{equation}
\Pr(T'_{i}<t)=1-\Pi_{j=1}^{r}\Pr(T_{i,j}\geq t) = 1-e^{-r\big(\mu \left( \frac{t}{2Cs\log{(2s)}}-1\right) \big)^\alpha}\  \forall\  t \ \geq  2Cs\log{(2s)}
\end{equation}

Let us denote $\tilde{F}_s(t)= \Pr(T'_{i}<t) $. Now, for repetition the probability of failure to meet a deadline occurs when at least one of the $P/r$ tasks have not completed. Thus the failure probability can be exactly computed as:
\begin{align}
P^f_s(t) &= \sum_{l=0}^{\frac{P}{r}-1} \binom{P/r}{l} (\tilde{F}_s(t))^l (1-\tilde{F}_s(t))^{P-l}  \\ 
 &= \sum_{l=0}^{\frac{P}{r}-1} \binom{P/r}{l} \Big(\frac{(\tilde{F}_s(t))}{1-\tilde{F}_s(t)}\Big)^l (1-\tilde{F}_s(t))^{\frac{P}{r}} \\
 &=\Theta \Big(c(P,K) \Big(\frac{\tilde{F}_s(t)}{1-\tilde{F}_s(t)}\Big)^{\frac{P}{r}-1} (1-\tilde{F}_s(t))^{\frac{P}{r}} \Big)
\end{align}
Here $c(P,K)$ denotes a function of $P$ and $K(=P-r+1)$, that is independent of $t$ (Note that $K=P-r+1$ for repetition). The last line follows since for $t$ large enough, $\frac{(\tilde{F}_s(t))}{(1-\tilde{F}_s(t))} \gg 1$, and thus for the purpose of analysis of the failure exponent for large $t$, the binomial summation is of the same order as the largest term, as we show here:
\begin{multline}
 \binom{P/r}{P/r-1} \Big(\frac{(\tilde{F}_s(t))}{1-\tilde{F}_s(t)}\Big)^{\frac{P}{r}-1} (1-\tilde{F}_s(t))^{\frac{P}{r}} \leq \sum_{l=0}^{\frac{P}{r}-1} \binom{P/r}{l} \Big(\frac{(\tilde{F}_s(t))}{1-\tilde{F}_s(t)}\Big)^l (1-\tilde{F}_s(t))^{\frac{P}{r}}  \\ \leq \sum_{l=0}^{\frac{P}{r}-1} \binom{P/r}{l} \Big(\frac{(\tilde{F}_s(t))}{1-\tilde{F}_s(t)}\Big)^{\frac{P}{r}-1} (1-\tilde{F}_s(t))^{\frac{P}{r}}.
\end{multline}

This implies,
\begin{multline}
 c_1(P,K)\Big(\frac{(\tilde{F}_s(t))}{1-\tilde{F}_s(t)}\Big)^{\frac{P}{r}-1} (1-\tilde{F}_s(t))^{\frac{P}{r}}  \leq \sum_{l=0}^{\frac{P}{r}-1} \binom{\frac{P}{r}}{l} \Big(\frac{(\tilde{F}_s(t))}{1-\tilde{F}_s(t)}\Big)^l (1-\tilde{F}_s(t))^{\frac{P}{r}} \\ \leq c_2(P,K)\Big(\frac{(\tilde{F}_s(t))}{1-\tilde{F}_s(t)}\Big)^{\frac{P}{r}-1} (1-\tilde{F}_s(t))^{\frac{P}{r}}.
\end{multline}

Here $c_1(P,K) =\binom{P/r}{P/r-1} $, \textit{i.e.}, a function of $P$ and $K(=P-r+1)$ only and independent of $t$. Similarly, $c_2(P,K)= \sum_{l=0}^{\frac{P}{r}-1} \binom{P/r}{l}  $ is also a function of $P$ and $K(=P-r+1)$ only and independent of $t$.
Thus,
\begin{equation}
P^f_s(t) = 
 \Theta \Big(c(P,K) \Big(\frac{\tilde{F}_s(t)}{1-\tilde{F}_s(t)}\Big)^{\frac{P}{r}-1} (1-\tilde{F}_s(t))^{\frac{P}{r}} \Big).
\end{equation}

\noindent Bounding the leading co-efficient in  the failure exponent, 
\begin{align}
 \lim_{t \to \infty}  \frac{\log(P^f_s(t))}{t^\alpha} 
& = \lim_{t\to \infty} 
\frac{\log \Big(c(P,K) \big(\frac{\tilde{F}_s(t)}{1-\tilde{F}_s(t)} \big)^{\frac{P}{r}-1} \frac{(1-\tilde{F}_s(t))^{\frac{P}{r}}}{1}\Big)} {t^\alpha}   \nonumber \\
& = \lim_{t\to \infty}  \left(\frac{P}{r}-1 \right)\frac{\log \left( e^{r\mu^\alpha \left( \frac{t}{(2Cs)\log{(2s)}}-1\right)^\alpha }-1\right)}{t^\alpha} + \frac{P}{r} \frac{\log \left(e^{-r\mu^\alpha \left( \frac{t}{(2Cs)\log{(2s)}}-1\right)^\alpha }\right)}{t^\alpha} \nonumber \\
& \to -\frac{r\mu^\alpha}{((2Cs)\log{(2s)})^\alpha} \nonumber \\
&= -\frac{\big( P-K  +1\big)\mu^\alpha}{((2Cs)\log{(2s)})^\alpha}
\end{align}

This proves the theorem.
\end{IEEEproof}

\section{Heuristic analysis of exponents}
\textit{Statement:} There exists an $( N_1 , N_2, P, s ) $ Coded Convolution strategy that outperforms the naive uncoded strategy in terms of asymptotic probability of failure if  $P (\log{P})^2 = o \left( \log{\sqrt{\frac{N_1 N_2}{P}}} \right)$  and $\alpha < 2\sqrt{\frac{P N_1}{N_2}} \left( 1+\frac{1}{ \log{2\sqrt{\frac{N_1 N_2}{P}}}} \right)^{-1}$.

\textit{Justification:} Consider the function given by
\begin{align}
 E(s) &=\log{(-\epsilon(s))} \\ 
 &=  \log\left(\frac{Ps}{N_2} -\frac{N_1}{s}  +1\right)-\alpha \log (2s\log{(2s)}) -\log\frac{\mu^\alpha}{C^\alpha}
\end{align}

A higher value of $E(s)$ implies faster decay of the failure exponent. Note that $s$ only takes values in the range $\sqrt{\frac{N_1 N_2}{P}} \leq s \leq \min\{N_1, N_2\}$. Here $s=\sqrt{\frac{N_1 N_2}{P}}$ corresponds to the uncoded strategy, and any $s$ strictly greater than $\sqrt{\frac{N_1 N_2}{P}}$ lies in the coded regime.

Now, we compare the exponents for uncoded and coded strategies. 

Consider the derivative of $E(s)$ in this range, as given by, 
\begin{equation}
E'(s)= \frac{ \frac{P}{N_2} + \frac{N_1}{s^2}  }{\left(\frac{Ps}{N_2} -\frac{N_1}{s}  +1\right)} -\frac{\alpha}{s} - \frac{\alpha}{s\log{(2s)}}
\end{equation}

At $s=\sqrt{\frac{N_1 N_2}{P}}$ (Uncoded), the derivative is given by,
\begin{equation}
E'(s)|_{\sqrt{\frac{N_1 N_2}{P}}} = \frac{2P}{N_2}-\frac{\alpha\sqrt{P}}{\sqrt{N_1 N_2}}-\frac{\alpha\sqrt{P}}{\sqrt{N_1 N_2} \log{2\sqrt{\frac{N_1 N_2}{P}}}}
\end{equation}
If $E'(s)|_{\sqrt{\frac{N_1 N_2}{P}}}> 0$, then $E(s)$ is strictly increasing, as $s$ increases beyond $\sqrt{\frac{N_1 N_2}{P}}$, \textit{i.e.}, into the coded regime. Thus, $E(s)$ attains a maximum in the regime  $s > \sqrt{\frac{N_1 N_2}{P}}$, \textit{i.e.}, in the coded regime. Thus, it is sufficient that
\begin{equation}
\alpha  < 2\sqrt{\frac{PN_1}{N_2}} \left( 1+\frac{1}{ \log{2\sqrt{\frac{N_1 N_2}{P}}}} \right)^{-1}
\end{equation}
This implies,
\begin{equation}
E'(s)|_{\sqrt{\frac{N_1 N_2}{P}}}> 0
\end{equation}
Thus there exists an $s$ in the coded regime for which the asymptotic failure probability is lesser than that in the uncoded regime.

\begin{figure}[!t]
\centering
\includegraphics[height=4cm]{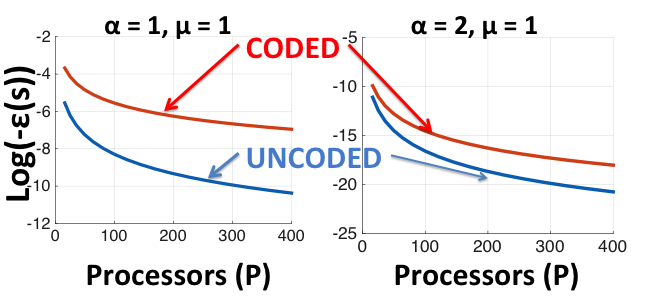}
\caption{Coding outperforms uncoded: Choose $N_1=4N$, $N_2=N$ and $P=\sqrt{\log{N}}$. Note that $s=\sqrt{\frac{N_1N_2}{P}}=\frac{2N}{\sqrt{P}}$ corresponds to the uncoded strategy, and for coded we choose $s= \frac{4N}{\sqrt{P}}$, strictly greater than uncoded. The plots show $\log{(-\epsilon(s))}$ for Uncoded( $s=\frac{2N}{\sqrt{P}}$) and Coded($s= \frac{4N}{\sqrt{P}}$) strategies as a function of $P$. }
\label{fig:s_path}
\end{figure}

Note that, $\alpha=1$, \textit{i.e.}, the shifted exponential also belongs to this regime. We also show some plots for the special case of $N_1=4N$, $N_2=N$ and $P=\sqrt{\log{N}}$, showing that choosing an $s$ in the coded regime outperforms uncoded strategy for the chosen values of $\alpha$ in Fig.~\ref{fig:s_path}. 
\end{document}